\documentclass[manuscript=preprint,layout=onecolumn,articletitle=true]{achemso}

\setkeys{acs}{usetitle = true}
\usepackage[version=3]{mhchem}
\usepackage{caption}
\usepackage{float}
\usepackage{geometry}
\usepackage{natbib}
\usepackage{setspace}
\usepackage{xkeyval}
\usepackage{color}
\usepackage[english]{babel}
\usepackage{epstopdf}

\newcommand{\onlinecite}[1]{\hspace{-1 ex} \nocite{#1}\citenum{#1}}
\newcommand{\subscript}[1]{\ensuremath{_{\textrm{\footnotesize  {#1}}}}}
\newcommand{\superscript}[1]{\ensuremath{^{\textrm{\footnotesize {#1}}}}}
\newcommand{\degree}{\ensuremath{^\circ}}

\author{A. Garcia-Lekue}
\affiliation[Donostia International Physics Center (DIPC)]{Donostia International Physics Center (DIPC)$,$ Paseo Manuel de Lardizabal 4$,$ E-20018 San Sebasti\'an$,$ Spain}
\alsoaffiliation{IKERBASQUE$,$ Basque Foundation for Science$,$ E-48011$,$ Bilbao$,$ Spain}
\email{wmbgalea@lg.ehu.es}

\author{M. Oll\'{e}}
\affiliation[Catalan Institute of Nanoscience and Nanotechnology (ICN2)]{Catalan Institute of Nanoscience and Nanotechnology (ICN2)$,$ UAB Campus$,$ E-08193 Bellaterra (Barcelona)$,$ Spain}
\alsoaffiliation[Eidgen\"{o}ssische Technische Hochschule (ETH) Zurich]{Department of Materials$,$ Eidgen\"{o}ssische Technische Hochschule (ETH) Zurich$,$ H\"{o}nggerbergring 64$,$ CH-8093 Zurich$,$ Switzerland}
\email{marcolle@gmail.com}

\author{D. Sanchez-Portal}
\affiliation[Donostia International Physics Center (DIPC)]{Donostia International Physics Center (DIPC)$,$ Paseo Manuel de Lardizabal 4$,$ E-20018 San Sebasti\'an$,$ Spain}
\alsoaffiliation[Centro de F\'{\i}sica de Materiales CSIC-UPV/EHU (CFM) ]{Centro 
de F\'{\i}sica de Materiales 
CSIC-UPV/EHU$,$ Paseo Manuel de Lardizabal 5$,$ San Sebasti\'an$,$ Spain}

\author{J.J. Palacios}
\affiliation[Dpto. de F{\'i}sica de la Materia Condensada, Univ. Aut{\'o}noma de Madrid]{Departamento de F{\'i}sica de la Materia Condensada,
 Universidad
 Aut{\'o}noma de Madrid, Cantoblanco, Madrid 28049, Spain}
\alsoaffiliation[Instituto de F{\'i}sica de la Materia Condensada (IFIMAC), Univ. Aut{\'o}noma de Madrid]
{ Instituto de F{\'i}sica de la Materia Condensada (IFIMAC), Universidad
 Aut{\'o}noma de Madrid, Cantoblanco, Madrid 28049, Spain}

\author{A. Mugarza}
\affiliation[Catalan Institute of Nanoscience and Nanotechnology (ICN2)]{Catalan Institute of Nanoscience and Nanotechnology (ICN2)$,$ UAB Campus$,$ E-08193 Bellaterra (Barcelona)$,$ Spain}

\author{G. Ceballos}
\affiliation[Catalan Institute of Nanoscience and Nanotechnology (ICN2)]{Catalan Institute of Nanoscience and Nanotechnology (ICN2)$,$ UAB Campus$,$ E-08193 Bellaterra (Barcelona)$,$ Spain}

\author{P. Gambardella}
\affiliation[Catalan Institute of Nanoscience and Nanotechnology (ICN2)]{Catalan Institute of Nanoscience and Nanotechnology (ICN2)$,$ UAB Campus$,$ E-08193 Bellaterra (Barcelona)$,$ Spain}
\alsoaffiliation[Eidgen\"{o}ssische Technische Hochschule (ETH) Zurich]{Department of Materials$,$ Eidgen\"{o}ssische Technische Hochschule (ETH) Zurich$,$ H\"{o}nggerbergring 64$,$ CH-8093 Zurich$,$ Switzerland}
\alsoaffiliation[Instituci{\'o} Catalana de Recerca i Estudis Avancats (ICREA)]{Instituci{\'o} Catalana de Recerca i Estudis Avancats (ICREA)$,$ E-08193 Barcelona$,$ Spain}

\title{Substrate-Induced Stabilization and Reconstruction of Zigzag Edges in Graphene Nanoislands on Ni(111)}


\begin{document}

%
%
%
%

\begin{abstract}
We combine experimental observations by scanning tunneling microscopy (STM) and density functional theory (DFT) to reveal the most stable edge structures of graphene on Ni(111) as well as the role of 
stacking-driven  activation and suppression of edge reconstruction.
Depending on the position of the outermost carbon atoms relative to hollow and
 on-top Ni sites, zigzag edges
have very different 
energies.
Triangular graphene nanoislands are exclusively bound by the more stable zigzag hollow edges. In hexagonal nanoislands, which are constrained by geometry to alternate zigzag hollow and zigzag top edges along their perimeter, only the hollow edge is stable whereas the top edges spontaneously reconstruct into the (57) pentagon-heptagon structure.
 Atomically-resolved STM images are consistent with either top-fcc
 or top-hcp epitaxial stacking of graphene and Ni sites, with the former being favored by DFT. 
Finally, we  find that there is a one-to-one relationship between the edge type, graphene stacking, and orientation of the graphene islands.
\end{abstract}

\section{Introduction}
Edges play a fundamental role in shaping the morphology\cite{loginova2009,wofford2010,Luo_acsnano11,artyukhov_PNAS12,ma2013} and electronic\cite{Nakada_prb96, Brey_prb06_2, Barone_nanolett06, ritter_nature09} properties of graphene nanostructures. Electron confinement due to edge boundaries gives rise to energy band gaps,\cite{Son_prl06,Han_PRL07} localized states\cite{Nakada_prb96,Brey_prb06_2}, spin-polarization,\cite{Son_prl06,Fernandez-Rossier_prl2007} and spin-dependent electron scattering.\cite{garcialekue_prl14} Furthermore, graphene edges determine the preferred sites for the attachment of metal atoms,\cite{Gan2008} chemical functionalization,\cite{Cervantes2008} as well as oxygen etching and intercalation.\cite{wang2010,Granas2012} Understanding and defining the edge morphology is therefore important to tune the growth of graphene nanostructures as well as to modulate the electronic properties of graphene in confined geometries.

In free-standing graphene, crystallographically oriented edges are of either zigzag (zz) or armchair (ac) type.\cite{Nakada_prb96}
Graphene with ac and zz edges can be obtained by mechanical exfoliation\cite{novoselov_science04} and etching techniques.\cite{Datta_nanolett08} However, free-standing unpassivated edges are unstable due to the high density of dangling bonds,\cite{koskinen_PRL08,mauri_PRL08,Huang_prl09} which induce the zz edges to spontaneously reconstruct into a line of pentagon and heptagon pairs, the so called zz(57) or Stone-Wales reconstruction.\cite{girit_science09, Koskinen_prb09}. A very different scenario arises in epitaxially grown graphene, where the interaction with the substrate can stabilize ac,\cite{Rutter_prb10, ChenAN2013} zz,\cite{LeichtAN2014, PharkPRB2012, LiAM2013} and  reconstructed edges,\cite{Prezzi_acsnano14, gao_JACS12, artyukhov_PNAS12, Tetlow2014pr} and induce complex graphene-metal boundaries.\cite{MerinoAN2014, PharkPRB2012, LiAM2013} 

Despite the potential of epitaxy for tailoring graphene edges evidenced by these studies, predicting the edge structure is still a challenging task, which requires detailed insight into the interplay of substrate interaction and edge morphology. For example, minimal variations in the lattice constant and electronic structure of the substrate can lead to very different edge energetics, resulting in hexagonal islands with zz and Klein-type edges in Co(1000),\cite{Prezzi_acsnano14} and either hexagonal or triangular islands in Ni(111), depending on growth parameters.\cite{Li13prb,olle_nanolett12} The latter strongly differ from the quasi-isotropic shape predicted by theory.\cite{gao_JACS12, artyukhov_PNAS12} 


In this work, we show that the experimentally observed morphological transition of graphene islands on Ni(111) is driven by a substrate-induced reconstruction of zz edges into the zz(57) structure.This reconstruction, which was found to be unstable on most close-packed metals including Ni(111) according to previous first principles calculations,\cite{gao_JACS12, artyukhov_PNAS12}, occurs along a particular direction with respect to the substrate, and can only be properly reproduced by theory by taking into account the stacking of differently oriented edges.

\section{Methods}

\subsection{Experiments}

Graphene nanoislands were grown on a Ni(111) single-crystal surface kept in a ultra-high vacuum (UHV) chamber with a base pressure of $3\times10^{-10}$ mbar. The crystal surface was cleaned by repeated cycles of Ar+ sputtering followed by annealing at 800~\degree C during 1 minute. The surface temperature was measured throughout the experiment using a pyrometer (IMPAC IGA 140). Propene was dosed on a freshly prepared Ni(111) crystal at room temperature. Once the dosing process is complete the sample is heated at 500~\degree C for 5 minutes to nucleate the graphene nanoislands and subsequently cooled to room temperature. The islands so obtained have an irregular shape. In order to prepare
triangular and hexagonal graphene islands (TGIs and HGIs, respectively) we post-anneal the sample to a temperature $T_A=500$~\degree C (650~\degree C) during a time $t_A=20$\,min (10\,min). We use a propene dose $D = 1$~L for the TGIs and $D = 2$~L for the HGIs to compensate for C loss at high temperature. The heating rate was 12\degree C/s and the cooling rate was 1.7\degree C/s.
  It should be emphasized that the presence of contaminants,
such as H or CO, is minimized by annealing the samples in UHV. Atomic H as well as molecular species, such as CO, desorb from Ni(111) at temperatures lower than 300~\degree C.\cite{IbachSS1980,ChristmannTJoCP1979} Moreover, we observe that the island shape does not change at 500\degree C, 
which  is another  indication  that the edge structures are not driven by hydrogenation.
Although it is notoriously difficult to completely rule out the
presence of contaminants, the fact that DFT calculations provide a very 
satisfactory explanation for the STM observations can be considered as a further
evidence for the absence of contaminants at the edges.
More details on the preparation of the TGIs and HGIs preparation can be found in Ref.\,\onlinecite{olle_nanolett12}. 
Topographic images of the surface were obtained at room temperature using a variable-temperature scanning tunneling microscope (SPECS, STM Aarhus 150) and processed using a freeware software (WSxM 5.0 develop 4.1).\cite{horcas_RSI07}

\subsection{Theory}

 Ab-initio spin-polarized calculations were performed using DFT as implemented in the SIESTA~\cite{siesta}
and ANT.G codes.\cite{ant.g} For the description of TGIs using SIESTA, we considered a supercell made of a three Ni(111) layers (up to four layers in convergence test, see the Supporting Information) and one graphene island containing 22 C atoms, whereas the results using ANT.G were obtained
employing a graphene island of 33 C atoms placed on
a Ni(111) surface described by a cluster with 2 Ni layers. The simulation of
 HGIs with alternate edge types would require the use of extremely large supercells; we have therefore used the nanoribbon geometry to compare the properties of different graphene edges on Ni(111) using SIESTA. The results obtained for the nanoribbon calculations yield information on the edge stability and energetics, which can be extended to the case of large hexagonal nanoislands. For such calculations we employed a 4x8 supercell made of three Ni(111) layers and a graphene ribbon containing 40 C atoms placed on one of the surfaces. In this case, only graphene nanoribbons with top-fcc stacking with respect to the underlying Ni(111) were considered. Further details on the calculations are given in the Supporting Information.

\section{Results and discussion}

 \begin{figure}[t!]
 \includegraphics[width=0.95\columnwidth]{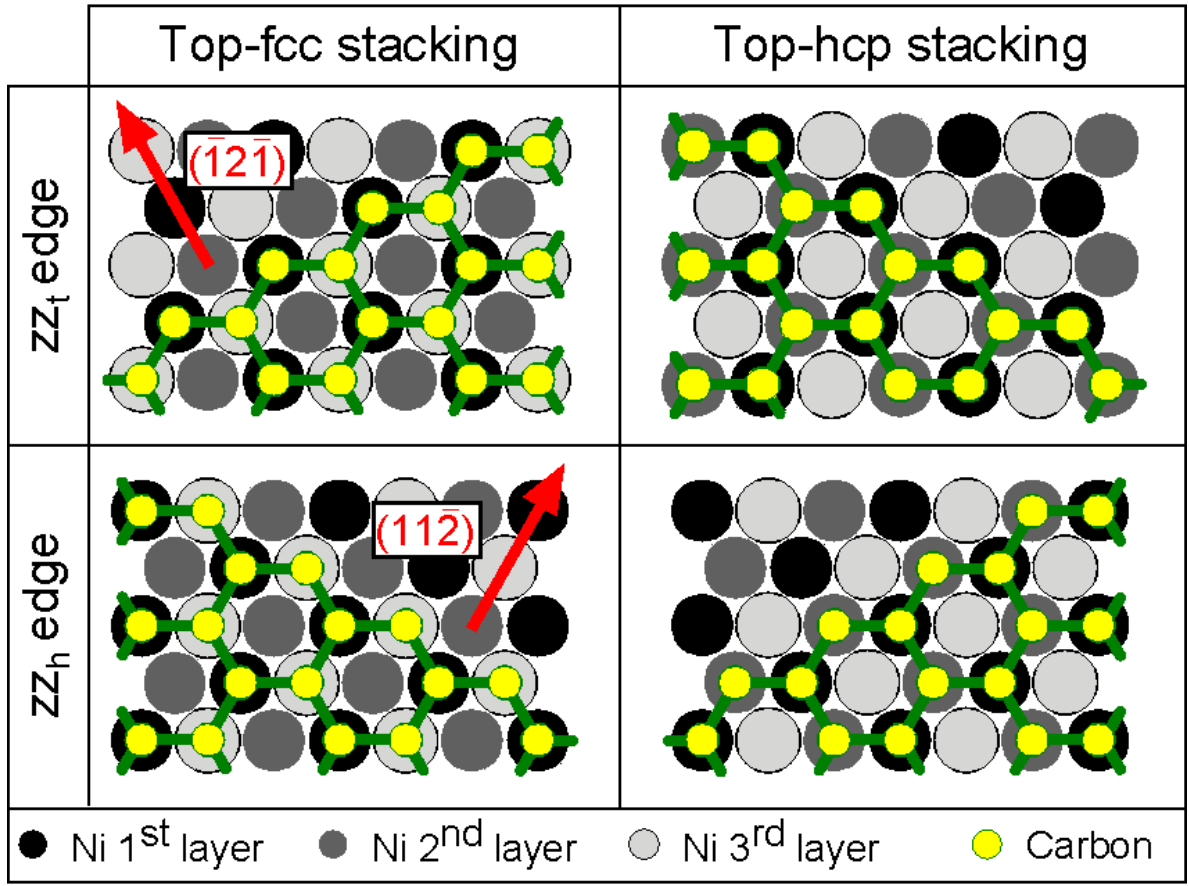}
 \caption{\label{fig1}
Hard sphere models of different zigzag graphene edges on Ni(111)
for fcc (left) and hcp (right) graphene-Ni stacking.}
 \end{figure}

The small lattice mismatch between graphene and Ni(111), 1$\times$1 stacking, and
the corresponding absence of a Moir\'e pattern make this system an optimal candidate to investigate the structure, stability, and epitaxial relationship between the graphene edges and a metallic substrate.
Several 1$\times$1 stacking structures have been proposed for graphene monolayers on Ni(111).\cite{gamo_SS97,
fuentescabrera_prb08,lahiri_nature10,zhao_JPCL11, kozlov_JPCC12}
According to these previous studies,
the top-fcc, top-hcp and bridge-top configurations
are considered to be the most stable, with a small energetic preference for the top-fcc stacking.
Indeed, a recent report  has confirmed the coexistance of
these three graphene configurations on Ni(111), with a general predominance of top-fcc.\cite{bianchini_JPCL14}  
In the case of graphene nanoislands on Ni(111) reported here, the possible stacking configurations are expected to be the same as for the graphene monolayer.
 Symmetry arguments, however, allow us to exclude bridge-top stacking (see
 Supporting Information) and consider only top-fcc and top-hcp stacking 
shown in Figure~\ref{fig1}.

Regarding zz edges, we distinguish two different types according to the position of the outermost C atoms with respect to the substrate. We thus define edges having the outer C atoms located on top of Ni atoms as zz\subscript{t} and edges having the outer C atoms on hollow sites as zz\subscript{h}, as shown in Figure~\ref{fig1}. The outer C atoms of a zz\subscript{h} edge can occupy hollow fcc or hcp sites, depending on the stacking configuration and edge orientation. As seen in Figure~\ref{fig1}, the crystallographic directions of zz\subscript{t} and zz\subscript{h} edges with the same stacking differ by 
60\degree\ and the crystallographic directions of edges of the same type but different stacking also differ by 60\degree. Note also that an edge perpendicular to the ($\overline{1}2\overline{1}$) direction can be either a top-fcc/zz\subscript{t} or a top-hcp/zz\subscript{h} edge, while an edge perpendicular to the ($11\overline{2}$) direction can be either a top-fcc/zz\subscript{h} or a top-hcp/zz\subscript{t} edge.

In previous work, we have shown that graphene islands that nucleate with random shape upon dosing propene on a clean Ni(111) surface and annealing the substrate above 450~\degree C evolve to either triangular or hexagonal shapes depending on the initial hydrocarbon dose, annealing time, and temperature.\cite{olle_nanolett12} Such TGIs and HGIs provide an ideal system to study the structure and stacking of the edge C atoms on close-packed metal surfaces. Due to the threefold symmetry of the zz\subscript{t} and zz\subscript{h} edges introduced above, TGIs present only a single edge type whereas HGIs must alternate different edges. By varying the island size and shape it is thus possible not only to characterize the different edges but also to study the influence of the edge morphology and graphene stacking on the stability of the islands.

\begin{figure}[t!]
\includegraphics[width=0.85 \columnwidth]{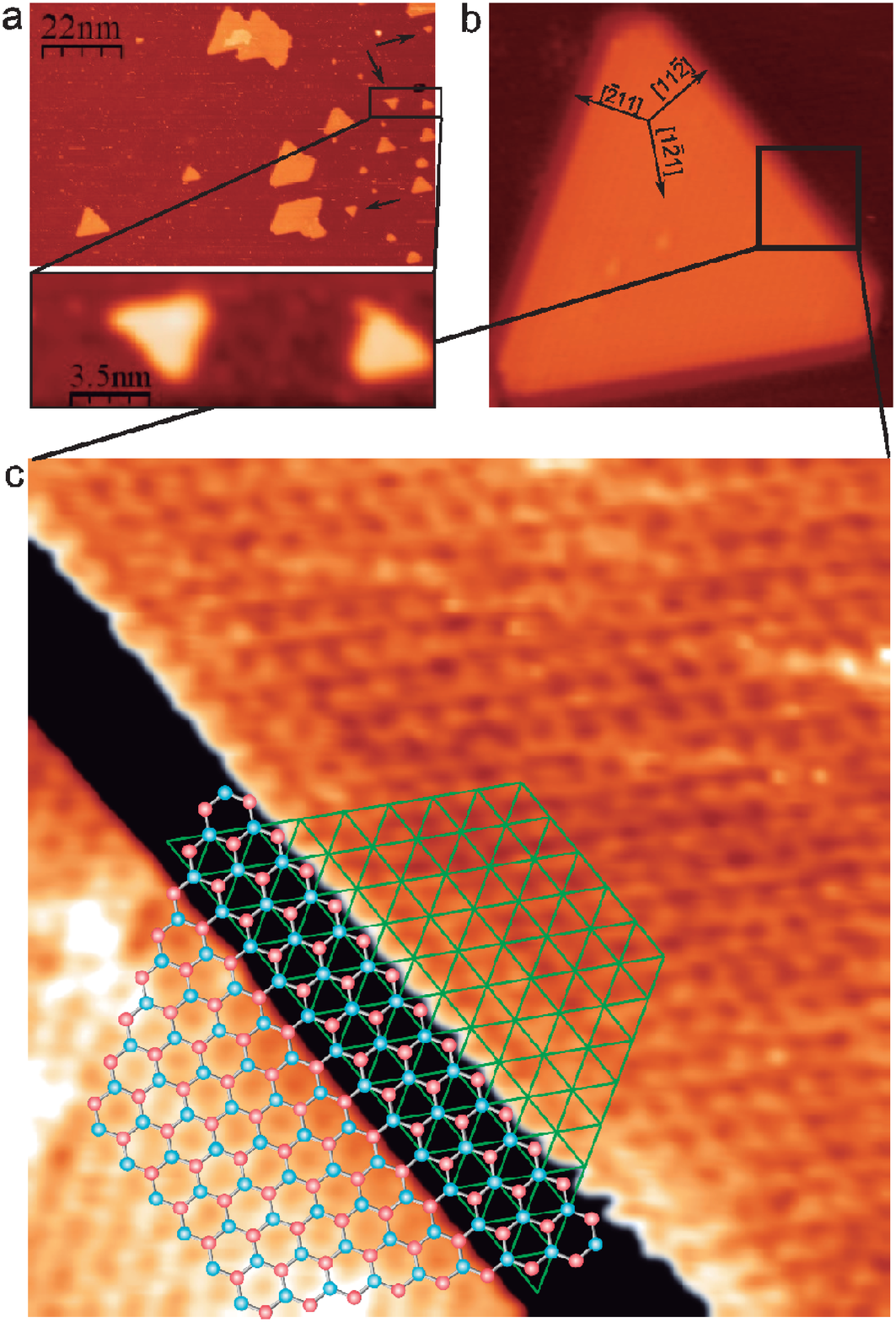}
\caption{\label{fig2}
(a) STM image of a large sample area with TGIs of different sizes. The enlarged image shows two small TGIs oriented in opposite directions. Setpoint current and bias voltage: $I = 1.9$ nA, $V_b = 3.7$ mV. (b) STM image of a single TGI. Setpoint: $I = 1.0$ nA, $V_b = 0.9$ mV. (c) Zoom-in of an edge region with the image contrast set to show the atomic resolution on both graphene and Ni layers.
The left part of the image corresponds to the graphene edge with a superposed honeycomb lattice (red and blue circles), the right part to the Ni(111) surface with a superposed hexagonal lattice (green mesh). The black stripe is a transition region related to the finite size of the tip where atomic resolution is missing.
}
\end{figure}

Figure~\ref{fig2}(a) shows a representative STM image of TGIs obtained by annealing the substrate to 500~\degree C and subsequent cooling to room temperature (see Methods). Under these conditions, most islands exhibit
straight edges, which indicates that they can only be of either zz or ac type. Topographic images with atomic resolution on both graphene and Ni are shown in Figure~\ref{fig2}(b) and (c).
 We observe that the TGIs have unreconstructed zz
 edges with no defects, in agreement with previous results.\cite{olle_nanolett12,Li13prb} The edge shown in Figure~\ref{fig2}(c) runs perpendicular to the [$11\overline{2}$] direction, i.e., parallel to a high symmetry [$1\overline{1}0$] direction of the Ni lattice. A honeycomb and an hexagonal lattice, representing the atomic positions of the graphene and surface Ni atoms respectively, have been superposed on the image for illustrative purposes.
Note that the STM image taken over the graphene-covered region
exhibits almost hexagonal contrast, despite the honeycomb atomic structure
of graphene. This is a well-known electronic effect
induced by the graphene-metal interaction, which breaks the symmetry between C atoms in on-top and hollow positions.\cite{dzemiantsova_PRB11,VarykhalovPRX2012,bianchini_JPCL14} 
In particular, previous theoretical simulations concluded that in the case of graphene on Ni(111) the bright spots in the STM images at
zero or very small bias  correspond to C atoms located in hollow positions, \cite{dzemiantsova_PRB11, bianchini_JPCL14} as reproduced also by our calculations.
Besides, all the images shown in this work
have been acquired under well controlled tip conditions, which
allows us to exclude tip-induced contrast effects.
All this suggests that, away from local variations related, e.g., to
 surface impurities, lattice imperfections, and the proximity of the edge, the spots with larger intensity in the STM images correlate with C atoms in hollow positions.
 In any case, the hexagonal units of the honeycomb lattice can be clearly identified and extrapolated to the Ni lattice, which let us draw two important conclusions: i) that  vertices of the honeycomb lattice at the graphene-Ni lateral boundary (red circles) correspond to hollow positions  on the hexagonal lattice, and hence the edge is of zz\subscript{h} type;  ii) The vertices representing the other sublattice (blue circles) are located on top of the Ni atoms, and hence graphene has either top-fcc or top-hcp stacking with the substrate, in agreement with our initial hypothesis.

The great majority of the TGIs in our samples point
in the same direction, as seen in Figure~\ref{fig2}(a). Because there is a one-to-one relationship between the edge type, stacking, and orientation of TGIs (Figure~S2 of the Supporting Information), this implies that the TGIs have a preference for a unique edge type and stacking combination, which, according to our analysis, corresponds to zz\subscript{h} and either top-fcc or top-hcp. Moreover, we find evidence that the edge type dominates the preference for the epitaxial stacking of the inner C atoms in the islands. The enlarged area of Figure~\ref{fig2}(a) shows two TGIs pointing in opposite directions, meaning
that they are rotated with respect to each other by 60\degree.  This
inverted orientation can only be explained by a change in the stacking or in the edge type (see Figure~S2 of the Supporting Information). Note that all the TGIs
pointing in the direction opposite to the predominant one [indicated by arrows in Figure~\ref{fig2}(a)] have a small size, below 10 nm$^2$. Since the effect of the edge type on the system's energy
is higher for smaller islands, the most plausible scenario is that the TGIs
 with inverted orientation maintain the lowest energy edge configuration and change their stacking.

\begin{figure}[t!]
\includegraphics[width=0.95 \columnwidth]{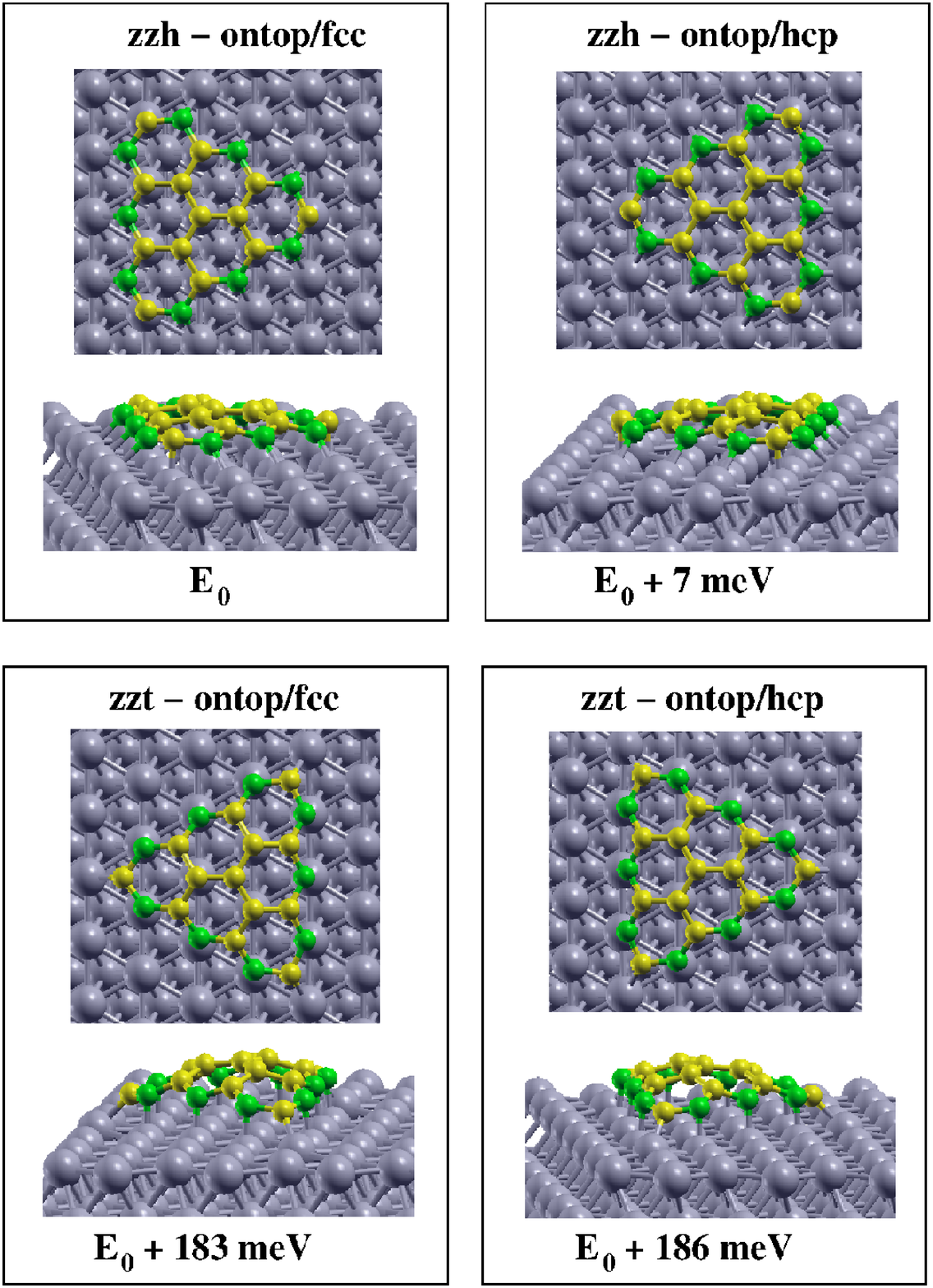}
\caption{\label{fig3}
 Relaxed model configurations of TGIs on Ni(111) calculated by DFT,
together with the corresponding energies per C atom in meV.
The latter are given with respect to E$_0$,
the energy of the most stable TGI.
}
\end{figure}

These experimental observations have been confirmed and complemented by DFT calculations. Figure~\ref{fig3} shows
the model systems used in the simulations, namely TGIs with zz\subscript{t} and zz\subscript{h} edges and top-fcc or top-hcp stacking. The TGI and the top two Ni layers of a three layer Ni slab are fully relaxed, which leads to significant plastic deformation of the islands owing to edge-induced relaxations. These effects are expected to be significant in small size islands such as the one used here, where the size is imposed by computational limitations. However, results obtained on larger islands using a
slab that contains only two metal planes confirm the edge energies obtained for the smaller TGIs (see
 Supporting Information). The relative energies of different edge types are calculated by subtracting the total energy of the TGIs shown in Figure~\ref{fig3} and dividing by the number of C atoms. This method neglects the plastic deformation of the islands, which would be difficult to disentangle by itself and that we estimate to have smaller weight compared to edge effects. Our results indicate an energy difference of $\sim$\,0.67~eV between C-edge atoms in hollow and top positions. From this,
we estimate an energy difference between zz\subscript{h} and
zz\subscript{t} edges of $\Delta\,E_{zz}\sim\,0.27\text{eV/\AA}$ (see the Supporting
Information for more details). This value might be overestimated because different edges induce
different relaxations in the inner part of the islands. However, this effect should be
mitigated for the calculations of larger islands reported in the Supporting
Information. In the following, we
use the average value of all these calculations to set the difference
of the zz\subscript{h} and
zz\subscript{t} edge energies, while our maximum and minimum values
are used to associate an error bar with this result. According to
this estimation $\Delta\,E_{zz} = 0.22 \pm 0.05 $ eV/\AA. Therefore, DFT predicts that the most stable TGIs have zz\subscript{h} edges, in agreement with the experimental results. The preference for zz\subscript{h} edges can be understood by considering the deformation of the sp\superscript{2} hybridized orbitals of the edge C atoms as they bind to the surface Ni atoms. In the case of zz\subscript{h} edges only a small deformation is required to form a bond with a Ni atom, whereas for zz\subscript{t} edges a larger deformation takes place, which also leads to the bent edge structure that can be observed in Figure~\ref{fig3}.

With respect to the stacking with the substrate, the calculations predict a small energy gain for the top-fcc stacking with respect to the top-hcp,
of the order of a few tens of meV per C atom, which is not very significant in the case of the small islands considered in this study. This result agrees with the calculations for extended graphene on Ni(111).~\cite{lahiri_nature10} Moreover, it is consistent with the preferential orientation found for the TGIs. The small energy difference obtained for the two stacking configurations is reflected by the fact that islands with opposite orientation, although very rare, can still be observed when their size is small enough  [Figure~\ref{fig2}(a)].

\begin{figure*}[t!]
\includegraphics[width= 0.95\textwidth]{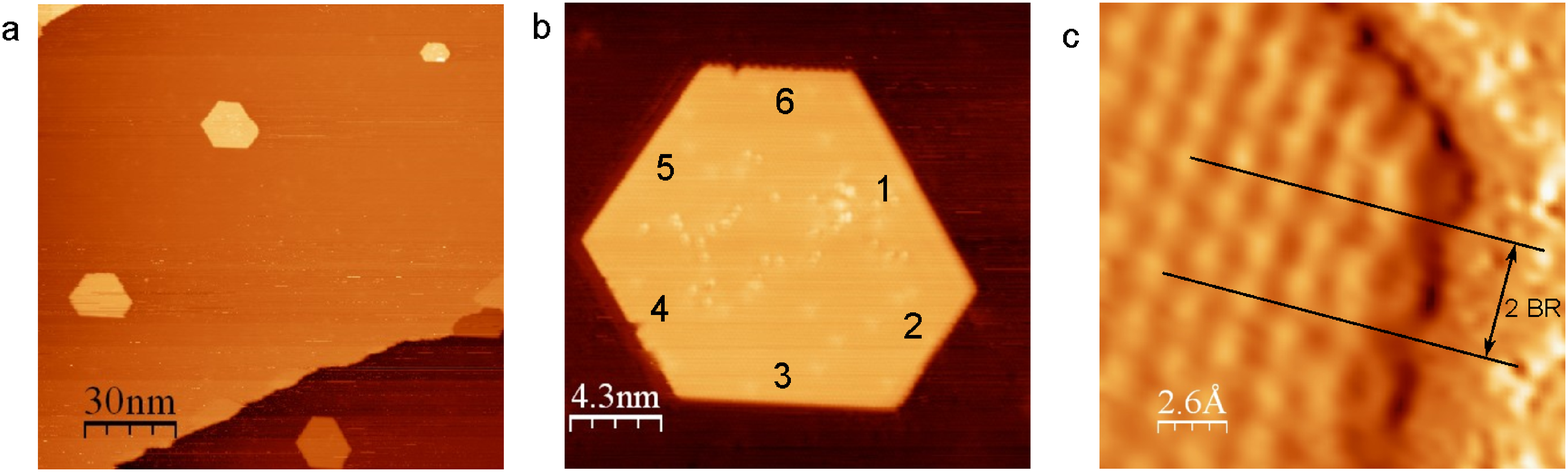}
\caption{\label{fig4}
(a) Large area STM image of HGIs.  Setpoint: $I = 1.9$ nA, $V_b = 2.1$ mV. (b) STM image of a single HGI.
Setpoint: $I = 9.6$ nA, $V_b = 0.9$ mV. (c) Detail of a corner of the HGI shown in (b) with zz\subscript{h} and zz\subscript{t}(57) edges. 
 The black lines in (c) illustrate  the doubling of the periodicity
 observed in the reconstructed edge. 
}
\end{figure*}

Figure~\ref{fig4}(a) shows that TGIs evolve into HGIs by annealing the Ni(111) surface up to 650~\degree C for 10 minutes. This seems to imply that the growth of the TGIs is kinetically limited, whereas the equilibrium shape of the graphene islands on Ni(111) is hexagonal, as confirmed by the results of our DFT
simulations. This is similar to the homo- and hetero-epitaxy of metals on (111) surfaces.\cite{MichelyPRL1993,PhysRevLett.81.1255} The triangular and hexagonal island shapes allow us to compare the structure and stability of different edge types. Contrary to TGIs, HGIs are constrained to exhibit zz\subscript{h} and zz\subscript{t} edges
alternated along their perimeter. Because of the 
60\degree\ angle between adjacent edges, the formation of ac edges can be excluded since these would be oriented at 
30\degree\ with respect to the zz edges. As in the case of TGIs, HGIs have their edges aligned with the high symmetry directions of the substrate, and hence possess only zz edges. Since zz\subscript{h} and zz\subscript{t} edges alternate independently of the stacking (Figure~\ref{fig1}), the edge energy contribution is equivalent for top-fcc
and top-hcp stacking. The orientation of the islands does not change as they evolve from triangular to hexagonal by increasing the annealing temperature from 500~\degree C to
650~\degree C, which indicates that the preferential stacking for the HGIs and TGIs is the same, that is, top-fcc according to the DFT results. Figure~\ref{fig4}(b) shows a detail of an HGI. Edges 1, 3, and 5 appear atomically straight and correspond to zz\subscript{h} edges. Edges 2, 4 and 6, on the other hand, present a few structural imperfections and are shorter compared to the odd-numbered edges. These edges should be of the zz\subscript{t} type according to the hard-sphere models presented in Figure~\ref{fig1}. However, atomic resolution images of the HGI show that these edges undergo a reconstruction that has a periodicity of two benzene rings. This is clearly visible in Figure~\ref{fig4}(c), where an island corner is shown at the intersection of edges 1 and 2. We assign this reconstruction to a pentagon-heptagon zz(57) structure, which has a periodicity of two benzene rings and
matches the observed atomic structure. 
 
\begin{figure}[h!]
\includegraphics[width=0.95\columnwidth]{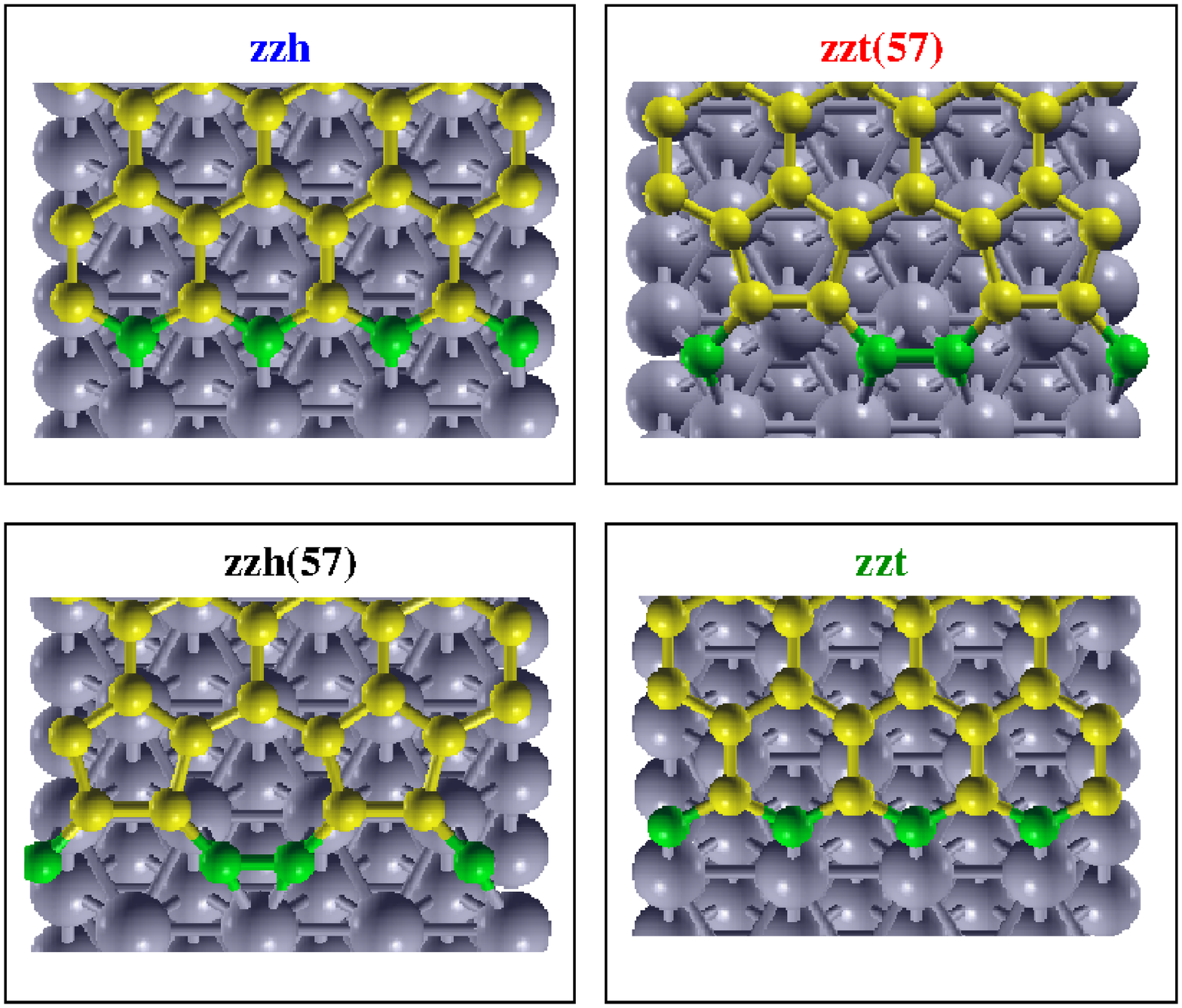}
\caption{\label{fig2_th}
Relaxed edge structures calculated for graphene nanoribbons with a width of 4 benzene rings on Ni(111).
For visualization purposes, the edge C atoms 
are represented in green.}
\end{figure}

In order to confirm these results, we have performed extensive DFT calculations of graphene nanoribbons on
Ni(111). Ribbons allow us to study two edge types simultaneoulsy, as HGIs, but are computationally less demanding compared to hexagonal islands.  In the following we only consider the zz(57) reconstruction, since the other known $2\times2$  reconstruction of a zz edge, labeled as zz(ad) in Ref.~\onlinecite{gao_JACS12}, is energetically excluded by calculations shown in the Supporting Information. Figure~\ref{fig2_th} presents the optimized structure of graphene  nanoribbons with zz\subscript{h}, zz\subscript{h}(57), zz\subscript{t} and zz\subscript{t}(57) edges on Ni(111). For such structures, as described in detail in the Supporting Information, we obtain the following relationships between the energy of the four edge types:
\begin{eqnarray}\label{eq2}
E_{zzt(57)}  = E_{zzt} - 0.15\,\text{eV/\AA} \nonumber \\
E_{zzh(57)} = E_{zzh} + 0.16\,\text{eV/\AA} \\
E_{zzh(57)} + E_{zzt(57)}  = 1.07\,\text{eV/\AA}. \nonumber
\end{eqnarray}
These numbers show that it is energetically favourable for zz\subscript{t} edges to undergo
the 57-reconstruction ($E_{zzt(57)}  < E_{zzt}$), whereas
the opposite is true for zz\subscript{h}  edges ($E_{zzh(57)} > E_{zzh}$), in agreement with the experimental observations.  

\begin{figure}[h!]
\includegraphics[width=0.95\columnwidth]{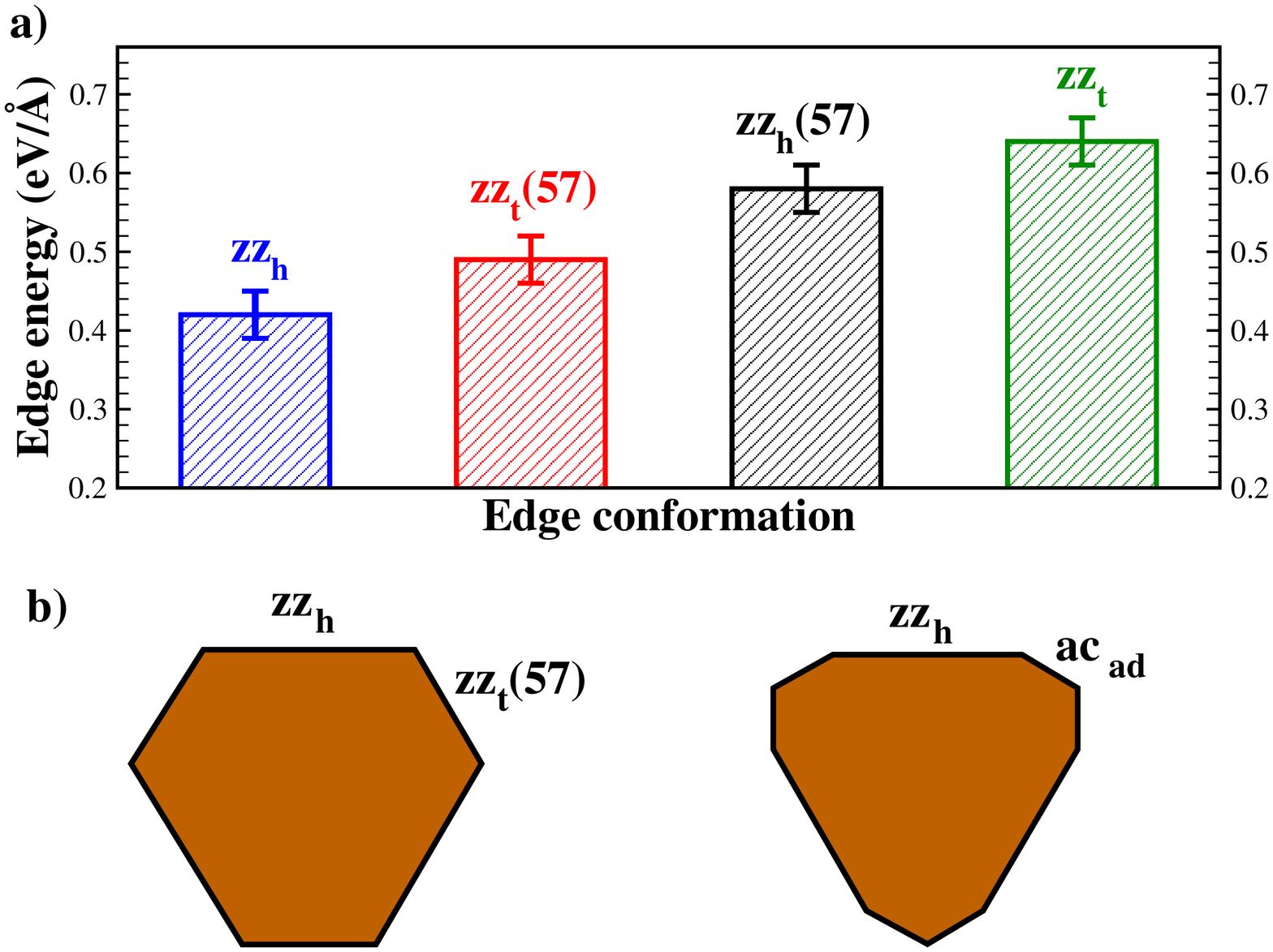}
\caption{\label{fig6}
(a) Formation energies of graphene zz edges on Ni(111).
(b) Equilibrium shape of graphene nanoislands on Ni(111) obtained by minimizing the edge energy. If edge reconstruction is allowed the islands exhibit zz$_h$ and reconstructed zz$_t$(57) edges, adopting an hexagon-like shape. If reconstruction is inhibited the islands have a triangle-like shape and
 exhibit zz$_h$ and reconstructed armchair edges.}
\end{figure}

From our calculations of the energetics of adsorbed nanoribbons
it is only possible to obtain, once stacking is taken into account,
three relations (Eq.~\plainref{eq2}) for four unknowns. Therefore,
additional information is required to estimate the
energy of the different edge types. Here we use our 
estimation 
of $\Delta\,E_{zz}$ obtained from the calculations of the TGIs.
In this way we obtain the edge energies represented
in Figure~\ref{fig6}(a), where the edges with lowest energy are
indeed the zz\subscript{h} and zz\subscript{t}(57) types.

The results of our calculations are in disagreement with
recent theoretical studies of graphene nanoribbons on Ni(111),\cite{gao_JACS12, artyukhov_PNAS12} in which
the unreconstructed edge is predicted to be more stable than the reconstructed one.
However, these studies do not consider the influence of stacking, which leads to the conclusion that the 57-reconstructed edges are always less stable than the unreconstructed ones.
Besides, at least in one of these works,~\cite{artyukhov_PNAS12}
the estimation of the edge formation energies is based on the assumption that the ribbons are perfectly symmetric, which is not true if epitaxial stacking is taken into account (see Figure~S4 of the Supporting Information). 
However, it is interesting to note that the average value (over the two stacking configurations)
of our edge energies is in good agreement with the values reported by 
Gao {\it et al.}~\cite{gao_JACS12} using plane-wave calculations and the same
exchange-correlation functional that we use here (Perdew, Burke and Ernzerhof functional~\cite{GGA}, see the Supporting Information). The agreement
is particularly good for the unreconstructed zigzag edges, for which our average
edge energy is 0.53~eV/\AA, exactly the value reported by Gao {\it et al.} for the
zigzag edge energy (without resolving the stacking dependency). 
Our average value for the 57-reconstructed edge energies is $\sim$0.54 eV/\AA, which 
is somewhat lower than the 0.60~eV/\AA\ reported by Gao {\it et al.}, but still
in good correspondence.

 With the edge energies in Figure~\ref{fig6}(a), following the classical Wulff construction, the equilibrium shape of graphene islands on Ni(111) turns out to be hexagonal, as shown in Figure~\ref{fig6}(b). The different length of the zz\subscript{h} and zz\subscript{t}(57) island edges observed in Figure~\ref{fig4}(b) 
can also be accounted for by such a model, 
and reflect the larger stability of zz$_h$ edges as compared
to zz$_t$(57) edges.
On the other hand, if no reconstruction would occur for the zz\subscript{t} edge,
the islands would have an almost triangular shape, as shown in Figure~\ref{fig6}(b) (the optimal
shape calculated for triangular islands exhibits also 
portions of reconstructed armchair instead of zz\subscript{t} edges, see the Supporting Information for more details).

These results outline a possible scenario to explain the
evolution of TGIs into HGIs on Ni(111) when the annealing temperature is increased from $\sim$500~\degree C to $\sim$650~\degree C. According to our theoretical model, HGIs with alternate zz\subscript{h} and zz$_t$(57) edges have lower energy compared to TGIs of equal size 
(we estimate HGIs of 10~nm$^2$ to be $\sim$3~eV more stable than
TGIs of the same size, see
the Supporting Information).  
This suggests that an energy barrier must be overcome
in order to achieve the equilibrium hexagonal shape. In view of our results, we can speculate that this energy barrier is associated with the 57-reconstruction of the zz edges.
Below a certain temperature the system does not have enough energy to overcome this barrier and, as a result, the zz\subscript{t} edge does not reconstruct and the island grows into a triangle.
On the contrary, as the temperature increases the barrier
can be easily surpassed and the zz\subscript{t} edges efficiently reconstruct into zz\subscript{t}(57), giving rise to hexagonal nanoislands.

\section{Conclusions}

In summary, we have shown that the structure and stability of the edges of graphene islands grown on Ni(111) are dominated by the stacking of the edge atoms relative to 
the 
substrate. Conversely, the edge type and energetics determine the island shape and, in TGIs smaller than 10~nm$^2$, also the stacking relationship of the inner C atoms with the substrate. Atomically-resolved STM images show that TGIs and HGIs exhibit only zz-like edges. As the epitaxial constraint imposed by the Ni substrate breaks the six-fold symmetry of free-standing graphene,
we distinguish between zz\subscript{h} and zz\subscript{t} edges, which differ in the position of the outermost C atoms relative to the substrate lattice. TGIs are bound uniquely by zz\subscript{h} edges, whereas HGIs are bound by alternate zz\subscript{h} and reconstructed zz\subscript{t}(57) edges. Accordingly, DFT calculations show that the energy of the zz\subscript{h} and zz\subscript{t}(57) edges is about 0.2~\text{eV/\AA} smaller relative to the unreconstructed zz\subscript{t} edges, which are not stable on this surface. The edge energetics fully accounts for the shape of the HGIs observed experimentally and suggests that the temperature driven transition from TGIs to HGIs is an activated process related to the existence of an energy barrier for the 57 pentagon-heptagon reconstruction. The stacking of the TGIs and HGIs is experimentally determined to be either top-fcc or top-hcp. All the TGIs and HGIs larger than a few nm$^2$ adopt the same stacking, which, according to DFT, corresponds to the top-fcc configuration. The above considerations are important when a well-defined $1\times1$ stacking leads to strongly orientation-dependent edge stacking configuration, as is the case of Ni(111) or Co(0001).\cite{Prezzi_acsnano14} This can apply
in a different degree also to other close-packed metal surfaces. Likewise, the stability of zz edges and their tendency to reconstruct may 
play a role in determining the density and orientation of grain boundaries in extended graphene layers grown on metal substrates.


\begin{acknowledgement}

We acknowledge support from the Basque Departamento
de Educaci\'on, UPV/EHU (Grant No. IT-756-13), the Spanish Ministerio de Ciencia e Innovaci\'on (Grants No.
 MAT2013-46593-C6-2-P and
MAT2013-46593-C6-5-P, MAT2007-62732), the ETORTEK program funded by the Basque Departamento de Industria,
the European Union
FP7-ICT Integrated Project PAMS under contract No. 610446,
 the European Research Council (StG 203239 NOMAD), and Ag\`{e}ncia de Gesti\'{o} d'Ajuts Universitaris i de Recerca (2014 SGR 715).
\end{acknowledgement}

\section{Supporting Information}
The atomic structures and symmetry of different edge types, TGIs, HGIs, as well as details of the DFT calculations can be found in the Supporting Information.
This information is available free of charge via the
Internet at http://pubs.acs.org


\providecommand{\latin}[1]{#1}
\providecommand*\mcitethebibliography{\thebibliography}
\csname @ifundefined\endcsname{endmcitethebibliography}
  {\let\endmcitethebibliography\endthebibliography}{}

\end{document}